 \documentstyle[12pt,twoside,fleqn,espcrc1]{article}

\newcommand{\AmS}{{\protect\the\textfont2
  A\kern-.1667em\lower.5ex\hbox{M}\kern-.125emS}}
\title{Theoretical Summary of the HADRON99 conference  }
\author{Harry J. Lipkin\address{Department of Particle Physics, 
        Weizmann Institute of Science, \\ 
        Rehovot, Israel }%
 \address{Israel Academy of Sciences and Humanities, \\ 
Jerusalem, Israel }%
 \address{School of Physics and Astronomy,
Raymond and Beverly Sackler Faculty of Exact Sciences,\\
        Tel Aviv University, Tel Aviv, Israel }%
\thanks  {Supported
in part by a grant from the United States-Israel
Binational Science Foundation (BSF), Jerusalem, Israel,
by the Basic Research Foundation administered by the Israel Academy of 
Sciences and Humanities}
      \address{High Energy Physics Division, Argonne National Laboratory,\\
Argonne, IL 60439-4815, USA}%
\thanks  {Supported
in part by the U.S. Department
of Energy, Division of High Energy Physics,  
Contract W-31-109-ENG-38.}}
\begin{document}
\maketitle
\begin{abstract}
The Constituent Quark Model has provided a remarkable description of the 
experimentally observed hadron spectrum but still has no firm theoretical 
basis. Attempts to provide a QCD justification discussed at Hadron99 include 
QCD Sum Rules, instantons, relativistic potential models and the lattice. 
Phenomenological analyses to clarify outstanding problems like the nature of
the scalar and pseudoscalar mesons and the low branching ratio for 
$\psi' \rightarrow \rho-\pi$ were presented. New experimental puzzles include
the observation of $\bar p p \rightarrow \phi \pi$.
\end{abstract}

\section{Introduction}

\centerline{Is there a theory?}

QCD is supposed to explain everything about Hadron Spectroscopy - But How?

QED is supposed to explain everything about Superconductivity - But How?

Will explaining Hadron Spectrocopy from first principles using QCD be as 
difficult as explaining Superconductivity from first principles using QED? 

\subsection{The Sakharov-Zeldovich 1966 Quark model (SZ66)}

Andrei Sakharov, a pioneer in quark-hadron physics asked in 1966
``Why are the $\Lambda$ and $\Sigma$ masses
different? They are made of the same quarks". Sakharov and 
Zeldovich\cite{SakhZel}. 
assumed a  quark model for hadrons with a flavor dependent linear mass
term and hyperfine interaction,
\begin{equation}
M = \sum_i m_i + \sum_{i>j} 
{{\vec{\sigma}_i\cdot\vec{\sigma}_j}\over{m_i\cdot m_j}}\cdot v^{hyp}_{ij} 
\end{equation}
where $m_i$ is the effective mass of quark $i$, $\vec{\sigma}_i$ is a quark 
spin operator and $v^{hyp}_{ij}$ is
a hyperfine interaction with different strengths
but the same flavor dependence
for $qq$ and $\bar q q$ interactions.

This model can be considered analogous to the BCS description of
superconductivity. The constituent quarks are quasiparticles of unknown
structure with a background of a condensate. They have effective masses not
simply related to the bare current quark masses, and somehow including all
effects of confinement and other flavor independent potentials. The only
contribution to hadron masses not already included is a flavor-dependent
two-body hyperfine interaction inversely proportional to the product of these
same effective quark masses. Hadron magnetic moments are described simply by
adding the contributions of the moments of these constituent quarks with Dirac
magnetic moments having a scale determined by the same  effective masses. The
model describes low-lying excitations of a complex system with remarkable
success. 

\subsection{Striking Results and Predictive Power}

Sakarov and Zeldovich already in 1966 obtained two relations between meson and
baryon masses in remarkable agreement with experiment.
Both the mass difference $m_s-m_u$ between strange and nonstrange quarks and
their mass ratio $m_s/m_u$ have the same values 
when calculated from baryon masses and meson masses\cite{SakhZel} 

\begin{equation}
\langle m_s-m_u \rangle_{Bar}= M_\Lambda-M_N=177\,{\rm MeV}
\end{equation}

\begin{equation}
\langle m_s-m_u \rangle_{mes} =
{{3(M_{K^{\scriptstyle *}}-M_\rho )
+M_K-M_\pi}\over 4} =180\,{\rm MeV}
\end{equation}

\begin{equation}
\left({{m_s}\over{m_u}}\right)_{Bar} =
{{M_\Delta - M_N}\over{M_{\Sigma^*} - M_\Sigma}} = 1.53 =
\left({{m_s}\over{m_u}}\right)_{Mes} =
{{M_\rho - M_\pi}\over{M_{K^*}-M_K}}= 1.61
\end{equation}

Further extension of this approach led to two more relations for $m_s-m_u$ when
calculated from baryon masses and meson masses\cite{ICHJLmass,HJLMASS}. and to
three magnetic moment predictions with no free parameters\cite{DGG,Protvino}

\begin{equation}
\langle m_s-m_u \rangle_{mes} =
{{3 M_\rho + M_\pi}\over 8}
\cdot
\left({{M_\rho - M_\pi}\over{M_{K^*}-M_K}} - 1 \right)
= 178\,{\rm MeV}.
\end{equation}
\begin{equation}
\langle m_s-m_u \rangle_{Bar}= 
{{M_N+M_\Delta}\over 6}\cdot
\left({{M_{\Delta}-M_N}\over
{M_{\Sigma^{\scriptstyle *}}-M_\Sigma}} - 1 \right)
=190\,{\rm MeV}.
\end{equation}
\begin{equation}
\mu_\Lambda=
-0.61
\,{\rm n.m.}=\mu_\Lambda =
-{\mu_p\over 3}\cdot {{m_u}\over{m_s}} =
-{\mu_p\over 3} {{M_{\Sigma^*} - M_\Sigma} \over{M_\Delta - M_N}}
=-0.61 \,{\rm n.m.}
\end{equation}

\begin{equation}
-1.46 =
{\mu_p \over \mu_n} =
-{3 \over 2}
\end{equation}

\begin{equation}
\mu_p+\mu_n= 0.88 \,{\rm n.m.}
={M_{\scriptstyle p}\over 3m_u}
={2M_{\scriptstyle p}\over M_N+M_\Delta}=0.865 \,{\rm n.m.}
\end{equation}

Also in 1966 Levin and Frankfurt\cite{LevFran} noted a remarkable systematics
in hadron-nucleon total cross sections indicating that mesons and baryons were
made of the same basic building blocks. The analysis supporting their ratio of
3/2 between baryon-nucleon and nucleon-nucleon cross sections has been 
refined\cite{LS} 
and consistently confirmed by new experiments\cite{PAQMREV}. Most recently the
new SELEX measurement\cite{SELEX}  $\sigma_{tot}(\Sigma p) = 36.96 \pm 0.65$ at
P =  609 GeV/c agrees with the prediction $\sigma_{tot}(\Sigma p) = 37.07$ mb.
from the 1975 two-component-Pomeron model (TCP)\cite{NewSys}, which also fit
all $\sigma_{tot}$, data now fit by PDG\cite{PDG} with fewer free parameters.
QCD calculations have not yet explained such remarkably successful simple
constituent quark model results. A search for new experimental input to guide
us is therefore of interest.

\subsection{How to go beyond SZ66 with QCD}

Many approaches are being investigated to use QCD in the description of hadron
spectroscopy. Some of these discussed at Hadron99 include: The lattice,
bag models, QCD Sum Rules\cite{Narison,Maltman}, Instantons\cite{Metsch},
Phenomenology\cite{Achasov,Bressani,Tuan}, Relativistic Potential
Models\cite{Metsch} and Reggeism and the Pomeron. Unfortunately there has been
no significant confirmation of any of these approaches by the kind of agreement
with experiment and the predictive power seen in the constituent quark model.
The complexity of QCD calculations necessitates the introduction of ad hoc
approximations and free parameters to obtain results, thus losing the simplicity
of the constituent quark model, with its ability to make many independent
predictions with very few parameters, There is also a tendency to lose some of
the good results of the constituent quark model; namely 

\begin{itemize}    
\item  The universal treatment of mesons and baryons made of the same quarks
\item  The spin dependence of hadron masses as a hyperfine interaction
\item  The appearance of the same effective quark masses in hadron masses, 
spin splittings and magnetic moments 
\item  The systematic regularities relating meson-nucleon and baryon-nucleon 
cross sections 
\end{itemize}

While none of these results can be considered to have a firm theoretical 
foundation based on QCD, it is difficult simply to dismiss the striking
agreement with experiment and the successful predictive power as purely 
purely accidental. 

The lattice approach, which seems to
have a firm foundation in QCD, has concentrated in attempting to fit known
properties of hadrons rather than predicting new hadron physics. In the one
area awaiting new exciting physics, the possible existence of glueballs, there
are ambiguous predictions and still no clear data. Hopefully these ambiguities
will be clarified by the next Hadron Spectroscopy conference. 

Underlying all the puzzles and paradoxes in hadron spectroscopy is the nature
of the pion, which seems to be at the same time a Goldstone boson and 2/3 of
a proton and a member of a pseudoscalar nonet whose masses are closer to the 
proton mass than to the pion mass. 

Present attempts to describe hadrons recall the story of the blind men and the
elephant\cite{PBIGSKY} . Each investigation finds one particular property of
hadrons and many contradictory conclusions arise,; e.g (1)  A pion is a
Goldstone Boson and a proton is a Skyrmion, (2) a pion is two-thirds of a
proton. The simple quark model prediction $ \sigma_{tot}(\pi^-p) \approx
(2/3)\cdot \sigma_{tot}(pp) $ \cite{LevFran,LS} still fits experimental data
better than 7\% up to 310 Gev/c\cite{PAQMREV}; (3) Mesons and Baryons are made
of the same quarks. Describing both as simple composites of asymptotically free
quasiparticles with a unique effective mass value predicts hadron masses,
magnetic moments and hyperfine splittings\cite{SakhZel,ICHJLmass,HJLMASS}. (4)
Lattice QCD can give all the answers, (5) Lattice calculations disagree on
whether the H dibaryon is bound and offer no hope of settling this question
until much bigger lattices are available\cite{PBIGSKY}. (5) Light (uds) SU(3)
symmetry and Heavy Quark symmetry (cbt) are good; (6) Light (uds) SU(3)
symmetry is bad. All nontrivial hadron states violate SU(3). All light V, A and
T mesons have good isospin symmetry with flavor mixing in (u.d) space and no $s
\bar s$ component; e.g. $\rho, \omega$. (7) The s-quark is a heavy quark.
Flavor mixing in mass eigenstates predicted by SU(3) is not there. Most
nontrivial strange hadron states satisfy (scb) heavy quark symmetry with no
flavor mixing.; e.g. $\phi, \psi, \Upsilon$. 

\subsection{Glueballs and Hybrids} 

The question of whether there are any hadron states which contain constituent 
gluons remains open. The main frontier here is experimental\cite{Chung}, 
with searches for
possible candidates described in detail at Hadron99. A theoretical review of
the status of glueballs and hybrids has been given by Narison\cite{Narison}

\section{Scalar and Pseudoscalar Mesons}

The lowest nine vector mesons are simply described as an ideal U(3) nonet with
states constructed from all possible combinations of (u,d,s) quark-antiquark
pairs. and with U(3) broken only by the difference between the masses of the
strange and nonstrange quarks. The lowest-lying pseudoscalar and scalar states
are very different and still present a confused picture in the hadron spectrum.

Most theoretical treatments treat scalars and pseudoscalars
very differently, with the pseudoscalar mixing arising from a peculiarly
pseudoscalar mechanism like PCAC, the low mass of the pion, and the U(1) 
problem. In contrast Metsch\cite{Metsch} treats both scalars and 
psudoscalars on the same footing and attributes the deviation from the ideal
vector spectrum to a single interaction due to instantons. Whether this can
provide a complete answer remains to be seen. 

New experimental input that can resolve many of the controversial issues can
be expected from investigations of heavy-quark hadron decays into final states
including these mesons\cite{sigmaE791,f0E791}.

\subsection{Problems with the $\sigma$, $a_o$ and $f_o$}

The scalar meson spectrum begins with three lowest lying states which do not
easily fit into any nonet: the $\sigma$, whose existence is still in question
and the $a_o$ and $f_o$. The general tendency here is to introduce other
configurations than $q \bar q$ for these states\cite{Maltman}. 

The question of whether the $a_o$ and $f_o$ mesons are $q \bar q$ states, 
$K \bar K$ molecules or four-quark states is still open and controversial.
All points of view were expressed at Hadron99. Narison\cite{Narison} using
QCD spectral sum rules  suggests that the $a_o$ is a $q \bar q$ state, while 
the $f_o$ is a quarkonium-gluonium mixture.  
Achasov\cite{Achasov} presents considerable phenomenological evidence for the 
four-quark nature 
of the $a_o$ and $f_o$. 
Metsch\cite{Metsch} describes the $f_o$ in the instanton model as a $q \bar q$ 
state in the lowest scalar nonet but has no place for the $a_o$.
Maltman\cite{Maltman} uses QCD Finite Energy Sum Rules to calculate the 
decay constants rather than the spectrum and suggests that scalar decay 
constants probe a spatial extent. 

Although no data on charmed meson decays into these controversial pseudoscalars
were presented at Hadron99, data presenting clear and extremely interesting
evidence for the existence and properties of the $\sigma$ as a light scalar
$\pi^+ \pi^-$ resonance observed in $D^+  -> \pi^+ \pi^- \pi^+$
decay\cite{sigmaE791} and for the masses and widths of $f_o$ observed in $D_s^+
-> \pi^+ \pi^- \pi^+$ decay\cite{f0E791} has been obtained at Fermilab. The
data should be available very soon, before the publication of these
proceedings.

\subsection{Problems with the $\eta$ and $\eta'$}

The pseudoscalars are conventionally decribed by adding an additional mass
contribution to the SU(3) singlet state, thus breaking U(3) while conserving
SU(3) and leaving SU(3) breaking as entirely due to quark mass differences. The
physical mesons $\eta$ and $\eta'$ are then described as mixtures described by
a mixing angle. either of either of the SU(3) singlet and octet states or of
the strange and nonstrange states analagous to the physical $\omega$ and $\phi$
vector mesons. Both mixing angles have been used at this
conference\cite{Narison,Metsch}. The dynamical origin of this additional
singlet contribution is still unclear and controversial, with some models
attributing it  the annihilation of an $q \bar q$ pair into gluons or
instantons\cite{Metsch}  and the creation of a $q \bar q$ pair of a different
flavor. Since the annihilation and creation processes are short-range and the
amplitudes are proportional to the initial and final state wave functions at
the origin there is no reason to limit the mixing to only ground state $q \bar
q$ wave functions and admixtures of radial excitations\cite{ICHJLpseud,Metsch}
and glueballs have been considered. 

\section{The Interface between Heavy Flavor Physics and Hadron Spectroscopy}

\subsection{Exotic Multiquark Hadrons - A Window Into QCD}

Two striking features of the hadron spectrum are (1) the absence of strongly
bound multiquark exotic states like a dipion with a mass less than two pion
masses or a dibaryon bound by 100 MeV. and (2) the description of nuclear
constituents as three-quark clusters called nucleons with no explicit quark 
degrees of freedom. The
constituent quark model gives a very simple explanation\cite{WhyNuc}. The one
gluon exchange ansatz for ${{V(q\bar q)_8}/{V(q\bar q)_1}}$ and ${{V(q
q)_6}/{V(q q)_{3*}}}$ gives: 

\begin{itemize}    
 
\item  Color-exchange color-electric interaction saturates\cite{TriEx} -
no forces between color singlet hadrons.
 
\item  Color electric energy unchanged by color recoupling.
 
\item  Color magnetic $qq$ forces repulsive for a single flavor - 
attractive between s quark in $D_s$ or $B_s$
and $u$ and $d$ quarks in proton.
 
\item  Energy gain by color-spin recoupling can bind
$H$ (hexaquark)\cite{Jaffe}, charmed and beauty 
pentaquarks\cite{PBIGSKY,Moriond,Houches,LipKEK,PANIC,Pentaqua,Gignoux} 
$P_c = \bar csuud$; $P_b = \bar bsuud$.
\end{itemize}    

The validity of this simple picture still remains to be confirmed by
experiment since no experimental information is yet available about
short-range color-sextet or color-octet two-body interactions.
All constituent quark model successes with a two-body color-exchange
interaction\cite{ICHJLmass,HJLMASS,DGG,TriEx} and all hadron spectroscopy
without exotics including scattering depend only upon
$(\bar qq)_1$ and $(qq)_{3*}$ interactions. 

Since physically realizable beams contain at least one $u$ or $d$ quark or
antiquark, hadron-nucleon scattering in the (u,d,s) sector is dominated either
by resonances produced by $\bar qq$ annihilation or by the repulsive
color-magnetic interaction keeping apart two hadrons containing quarks of the
same flavor. Only with more than three flavors can the $(qq)_6$ or $(\bar
qq)_8$ interactions be observed in realistic scattering experiments with no
common flavor between beam and target. Thus the possible existence of exotic
hadrons remains crucial to understanding how QCD makes hadrons from quarks and
gluons\cite{Lipk86}. 

The $H$ dibaryon\cite{Jaffe} was shown to have a gain in color-magnetic 
energy over the $\Lambda \Lambda$ system\cite{Jaffe,Rosner}. But a lattice 
calculation\cite{Thacker} showed a repulsive $\Lambda$-$\Lambda$ interaction 
generated by quark exchange\cite{Pstanford,Thacker2} not included in 
simple model calculations which could well prevent the six quarks from coming 
close enough together to
feel the additional binding of the short range color-magnetic interaction.
Pentaquarks, shown
\cite{Moriond,Houches,LipKEK,PANIC,Pentaqua,Gignoux} to have a color-magnetic 
binding roughly equal to the 
$H$, have no possible quark exchange force in the lowest decay channel
$D_sN$ \cite{Pstanford}.
The simplest lattice calculation 
can easily be done in parallel with the more complicated H calculation
both in the symmetry limit where all light quarks have the same mass
and with $SU(3)$ symmetry breaking.
Comparing results may provide
considerable insight into the physics of QCD in
multiquark systems even if the pentaquark is not bound. 
However, no such lattice calculation has been done or is planned.

The experimental searches for the $H$ abd the pentaquark have been summarized
\cite{Asheryfour,Asherysix} with so far no conclusive evidence for either. A
few candidate pentaquark events have been reported\cite{Asheryfive,Asheryseven},
but the evidence that these are not due to systematic errors is not convincing.
A better approach would be to search with a good vertex detector and good
particle ID for secondary vertices where one of the particles emitted from the
secondary vertex is unambiguously a proton\cite{PBIGSKY}. This immediately
identifies the particle as a weakly decaying baryon and something new and
interesting if its mass is not equal to the mass of the known charmed or beauty
baryons.

\subsection{New $b \bar c$ Spectroscopy and possible tetraquarks}
   The new $b \bar c$ mesons found at Fermilab introduce a new field of 
hadron spectroscopy. In addition to the normal meson spectrum, there is the
question of $bcq$ baryons and $bc\bar u \bar d$ tetraquarks. The $bc$ diquark
can be produced in a high energy hadron experiment with cross sections 
comparable to the $b \bar c$ meson. The $bc$ diquark can then capture a light 
quark $q$ to produce a $bcq$ baryon. But the $bc$ diquark will have a much 
smaller size than diquarks containing lighter flavors and can appear on the
light quark mass scale as a point particle with the color quantum numbers of
an antiquark. The $bc$ diquark can therefore attract two light antiquarks to
produce a $bc\bar u \bar d$ tetraquark with a light quark wave function 
similar to that of an antibaryon containing one heavy antiquark.     

\subsection{Interface between QCD and Weak decays }

Are there pseudoscalar or scalar resonances near the $D$ and $D_s$ masses; e.g. 
the $\pi (1800)$? If so, how do they affect weak decays. There is room for both 
theoretical and experimental investigation. Many final state channels are open
in this mass range and similarities in branching ratios for strong and weak 
decays can provide interesting clues to the structure of these hadrons. In 
particular, there could be evidence for hybrid structure\cite{cll}.

The $B$ and $D$ decays to $\eta$, $\eta'$, $\sigma$, $f_o$ and $a_o$ can 
provide new and independent information regarding the structure of these 
controversial mesons. The decay $B \rightarrow K \eta'$ is much too large for 
any model, while $B \rightarrow K \eta$ is much smaller and not yet seen. 
However in the final states with a $K^*$ rather than a $K$, the decay 
$B \rightarrow K^* \eta$ 
has been seen and $B \rightarrow K^* \eta'$ has not.
The decay $D_s \rightarrow f_o \pi$ is strongly seen in the decays
$D_s \rightarrow 3 \pi$\cite{f0E791}. Will $D_s \rightarrow a_o \pi$ be seen in
$D_s \rightarrow \eta  \pi$?
The $D \rightarrow \sigma \pi$ has been seen in $D \rightarrow 
3 \pi$\cite{sigmaE791}. 

While the search for new evidence of CP violation provides the main motivation
for investigations of $B$ and $D$ decays, the interpretation of these decays 
requires some input from final state interactions and QCD. Thus there is another
connection between heavy flavor physics and hadron spectroscopy.

\subsection{Is the strange quark a heavy quark?}

 Most nontrivial strange hadron states
satisfy (scb) heavy quark symmetry with no flavor mixing.; 
e.g. $\phi, \psi, \Upsilon$. 
The strange axial mesons are particularly interesting. The triplet and singlet
quark spin states $^3P_1$ and $^1P_1$ which are SU(3) eigenstates are badly 
mixed. HQET couples the light quark spin to the orbital $L=1$ to get to doublets
with j=1/2 and j=3/2 instead of a triplet and a singlet..
Which is it?

\section{Open puzzles in hadron spectroscopy}

\subsection{Constituent Quarks vs. Current Quarks}

Hadron spectroscopy tends to decribe the known mesons and baryons as $q \bar q$
and $3 q$ systems. However, deep inelastic scattering experiments lead to a 
description of hadrons containing these states as valence quarks and adding 
a sea of quark-antiquark pairs and gluons\cite{Ma}. There seems to be a general
consensus that this picture should not be discussed in detail in a Hadron 
Spectroscopy conference. It is interesting hadron physics, but belongs 
elsewhere; e.g. in meetings on the spin structure of the nucleon, strangeness 
in the nucleon and deep inelastic scattering. However, a few points from this
picture did arise at the conference. 

\subsubsection{The observation of $\bar p n \rightarrow \phi \pi $; 
Strangeness in the proton?}

Does this observation\cite{Bressani} indicate strangeness in the nucleon or perhaps 
something more exciting?

If $\bar p n \rightarrow \phi \pi$, then $\phi \pi \rightarrow \bar p n$.
This would be the first experimental evidence for structure in $\phi \pi$
scattering in one partial wave. What are the dynamics of this scattering?
Does it go via a hybrid or some other exotic state that couples to $\phi \pi$?
What other final states are coupled to this partial wave in $\phi \pi$ 
scattering? More clever experiments are needed to unravel this mystery. 

\subsubsection{The spin structure of the $\Lambda$}

In the constituent quark model the $\Lambda$ contains two nonstrange quarks
whose spins are coupled to spin zero and the strange quark carries the entire
spin of the baryon. But in the picture where strange quarks carry some of the
spin of the proton, the nonstrange quarks carry some of the spin of the 
$\Lambda$. Possible experimental tests of this description were 
discussed\cite{Ma}.  

\subsection{The $\psi' \rightarrow \rho \pi$ dilemma}

That the experimental ratio of the branching ratios 
$BR(\psi' \rightarrow \rho \pi)/BR(J\psi' \rightarrow \rho \pi)$
is much smaller than the experimental values for ratios of decays of 
$\psi'$ and $J\psi$ into other decay modes is well known and so far defies 
all theretical explanations. An update of this problem was resented\cite{Tuan}.

\subsection{Heavy flavor decays to pseudoscalar mesons} 

The ratios of the $D_s$ branching ratios 
${{BR(D_s \rightarrow \eta' \pi)}/{BR(D_s \rightarrow \eta \pi)}}$
and 

\noindent ${{BR(D_s \rightarrow \eta' \rho)}/{BR(D_s \rightarrow \eta \rho)}}$ 
are both much too large to be explained by any conventional model.

\section{Conclusions}

There are many theoretical approaches to applying QCD to hadron spectroscopy. No
one seems far superior to the others. There are strong disagreements between 
various approaches to the description of the scalar and pseudoscalar mesons. New
data can help resolve these open questions.

Some open puzzles:

$\bar p p \rightarrow \phi \pi$ and $\phi \pi \rightarrow ?? $

$\psi' \rightarrow \rho \pi $ 

Interface with heavy flavor physics should be developed; e.g. $D$ and $B$ decays
to $f_o$, $a_o$, $\eta$ and $\eta'$.

Exotic tetraquarks $(bc \bar u \bar d)$ and pentaquarks $(\bar Qsuud)$ should be
investigated both experimentaly and theoretically.

Possible applications of HQET also considering the strange quark as a heavy
quark

\end{document}